%% file: main.tex
\documentclass[aps,physrev,10pt,floatfix,longbibliography,nofootinbib,reprint]{revtex4-2}
\usepackage{amssymb}
\usepackage[english]{babel}
\usepackage{bm}
\usepackage{graphicx}
\usepackage{hyperref}
\usepackage{physics}

\newcommand{\arr}[1]{{\bm{#1}}}
\newcommand{\bmsi}{\arr{\sigma}}
\newcommand{\calB}{\mathcal{B}}

\newcommand{\calH}{\mathcal{H}}

\newcommand{\diag}{\mathrm{diag}}
\newcommand{\rme}{\mathrm{e}}
\newcommand{\rmi}{\mathrm{i}}
\newcommand{\rmT}{\mathrm{T}}
\newcommand{\spindown}{{\downarrow}}
\newcommand{\spinup}{{\uparrow}}

\newcommand{\Eq}[1]{Eq.~(\ref{eq:#1})}
\newcommand{\Fig}[1]{Fig.~\ref{fig:#1}}

\begin{document}

\title{From Tensor Network Quantum States to Tensorial Recurrent Neural Networks}
\author{Dian Wu} \email{dian.wu@epfl.ch}
\author{Riccardo Rossi} \email{riccardo.rossi@epfl.ch}
\author{Filippo Vicentini} \email{filippo.vicentini@epfl.ch}
\author{Giuseppe Carleo} \email{giuseppe.carleo@epfl.ch}
\affiliation{Institute of Physics, \'Ecole Polytechnique F\'ed\'erale de Lausanne (EPFL), CH-1015 Lausanne, Switzerland}
\date{\today}

\begin{abstract}
We show that any matrix product state (MPS) can be exactly represented by a recurrent neural network (RNN) with a linear memory update.
We generalize this RNN architecture to 2D lattices using a multilinear memory update.
It supports perfect sampling and wave function evaluation in polynomial time, and can represent an area law of entanglement entropy.
Numerical evidence shows that it can encode the wave function using a bond dimension lower by orders of magnitude when compared to MPS, with an accuracy that can be systematically improved by increasing the bond dimension.
\end{abstract}

\maketitle

\section{Introduction}

Tensor networks (TN) have been extensively used to represent the states of quantum many-body physical systems~\cite{baxter1968dimers, affleck1987rigorous, takasaki1999fixed}. Matrix product states (MPS) are possibly the simplest family of TN, and are suitable to capture the ground state of 1D gapped Hamiltonians~\cite{verstraete2006matrix, hastings2007area}. They can be contracted in polynomial time to compute physical quantities exactly, and optimized by density matrix renormalization group (DMRG)~\cite{white1992density} when used as variational ansätze. More powerful TN architectures that cannot be efficiently contracted in general have been proposed later, notably projected entangled pair states (PEPS)~\cite{schuch2007computational}. They are of interest mainly because they can describe area-law states in dimensions $D > 1$~\cite{Lami22BackflowTN}. In practice, they are usually approximately contracted and optimized with low-rank truncation~\cite{evenbly2017algorithms, pan2020contracting, vanderstraeten2021variational}. They can also be optimized in the variational Monte Carlo (VMC) framework~\cite{sandvik2007variational, sfondrini2010simulating, liu2021accurate, vieijra2021direct} using stochastic sampling rather than exact contraction.

In recent years, a growing trend has been to use TN in machine learning tasks such as supervised~\cite{stoudenmire2016supervised, Cheng21PRBSupervisedTN} and unsupervised~\cite{Han18UnsupervisedMPS, Liu22UnsupTN} learning. It is of particular interest to use them as generative models~\cite{Cheng19TTNGen, Vieijra22GenPEPS}, where it is possible to draw perfect samples with tractable probability densities. Considering that many generative models are based on recurrent neural networks (RNN), there have been attempts to use tensor operations in RNN to improve their expressivity~\cite{sutskever2011generating, irsoy2014modeling}. This kind of architectures has shown leading performance when applied to quantum physical systems~\cite{hibat2021variational}.

Soon after the original proposal of neural quantum states (NQS)~\cite{carleo2017solving}, several attempts have been made to link them to ground states of well-known Hamiltonians and TN quantum states. While an arbitrary Hamiltonian might require an NQS with a depth growing exponentially with the system size~\cite{Carleo2018NatCommDeepRBM}, several Hamiltonians with desirable topological properties have compact representations in NQS~\cite{Deng17PRBNQSTopological, Glasser2018PRX, Kaubruegger18PRBChiralNQS, Lu19PRBTopologicalNQS}.

Considering the relation between neural networks (NN) and TN, the first works focused on the restricted Boltzmann machines (RBM), which are one of the simplest classes of NN.
It is impossible to efficiently map an RBM onto a TN, as they correspond to string-bond states with an arbitrary nonlocal geometry~\cite{Glasser2018PRX}.
This result was later refined to show that an RBM may correspond to an MPS with an exponentially large bond dimension, and only short-range RBM can be mapped onto efficiently computable entangled plaquette states~\cite{Chen18PRBEquivalence}.
Similar results have been obtained that deep Boltzmann machines with proper constraints can be mapped onto TN that are efficiently computable through transfer matrix methods~\cite{Pastori19PRB}.

Among other classes of NN, it has been shown that simple cases of convolutional neural networks (CNN) and RNN, which are also instances of arithmetic circuits or sum-product networks~\cite{poon2011sum}, can be efficiently mapped onto TN. Such efficient mapping is no longer possible with modifications to those simple cases, including overlapped convolution kernels, stacked recurrent layers~\cite{levine2019quantum}, or nonlocal score function~\cite{khrulkov2018expressive}, which suggests that the reuse of information in general NN architectures is an extensive source of expressivity. In the opposite direction, a mapping from arbitrary TN to feedforward NN can be constructed using NN layers that contract a set of tensor indices at a time~\cite{Sharir22NQSTN}.

In this work, we propose a new family of variational ansätze that shares characteristics with NN and TN. We start by establishing an exact mapping from MPS to an RNN with a linear update rule for the memory and a quadratic output layer, which we refer to as 1D~MPS-RNN. We then propose a generalization of this architecture to 2D lattices using a linear combination of memory components at previous neighboring sites in each memory update, which we call 2D~MPS-RNN. After a theoretical discussion of the limitation on entanglement entropy due to the linear memory update, we propose a multilinear update with inspiration from the tensorial nature of the PEPS architecture. This ansatz can represent states with area law of entanglement entropy, which is particularly interesting in quantum physics. Unlike PEPS, our ansatz is still a generative model, which supports perfect sampling and wave function evaluation in polynomial time. Finally, we conduct numerical experiments to compare the performances of MPS and our ansätze on the antiferromagnetic Heisenberg model (AFHM) on square lattice and triangular lattice. The results show that our ansätze produce lower variational energy with bond dimensions smaller by orders of magnitude compared to MPS, even in the presence of frustration, and the trade-off between computation cost and accuracy can be systematically controlled by the bond dimension, which is the only hyperparameter of our architectures.

\begin{figure}[t]
\includegraphics[width=\linewidth]{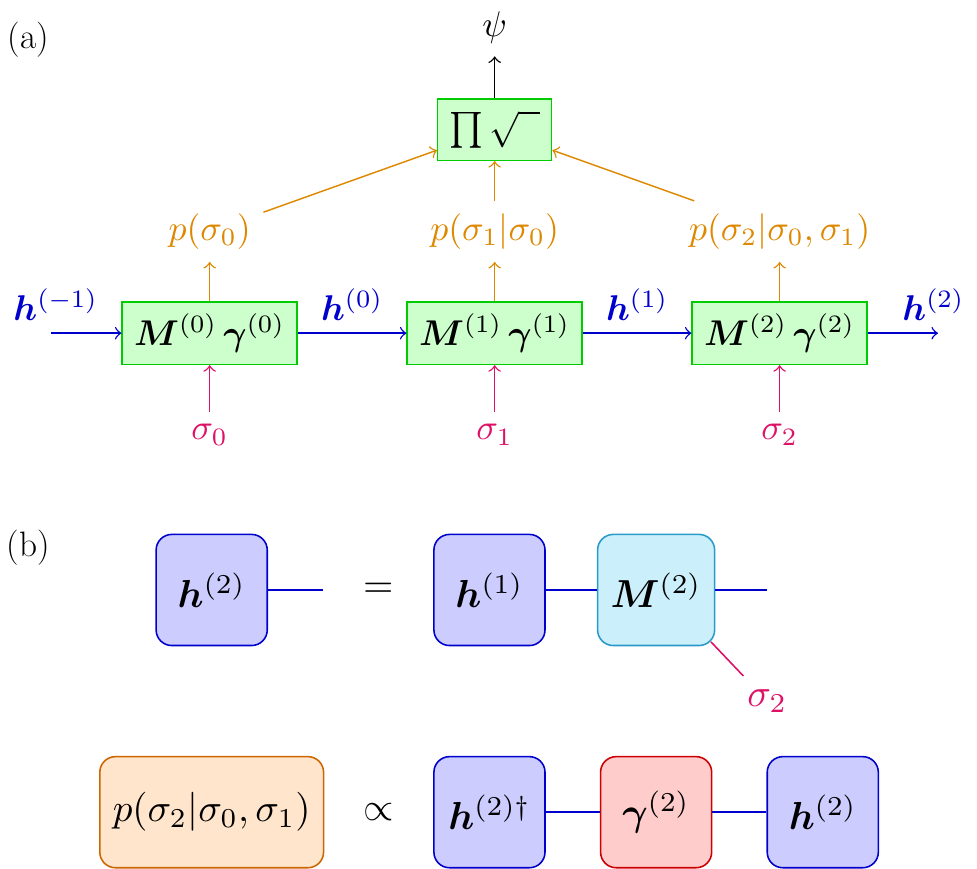}
\caption{
(a) Computational graph of the vanilla MPS-RNN quantum state. The phase factor $\phi(\bmsi)$ is omitted in the figure.
(b) TN diagrams of the memory $\arr{h}^{(i)}$ and the conditional probability $p(\sigma_i | \bmsi_{< i})$ for the vanilla MPS-RNN.
}
\label{fig:mps-rnn}
\end{figure}

\section{RNN quantum states}

We can write any wave function $\psi$ of a quantum spin-$1/2$ system consisting of $V$ sites in the form
\begin{equation}
\psi(\bmsi) = \left( \prod_i \sqrt{p(\sigma_i | \bmsi_{< i})} \right) \rme^{\rmi \phi(\bmsi)},
\label{eq:mps-rnn-psi}
\end{equation}
where $\bmsi = (\sigma_0, \ldots, \sigma_{V - 1})$ is the spin configuration in the $z$-basis, and $p(\sigma_i | \bmsi_{< i})$ is the conditional probability of measuring the spin $i$ given the measurements of previous spins $\bmsi_{< i} = (\sigma_0, \ldots, \sigma_{i - 1})$.
An RNN quantum state~\cite{hibat2020recurrent} uses a memory vector $\arr{h}^{(i)}$ at each step to summarize the information about the previous spins $\bmsi_{\le i}$, which is updated by a function $\arr{h}^{(i)} = f^{(i)}_\text{update}(\sigma_i, \arr{h}^{(i - 1)})$. Using this information, we compute each conditional probability from an output function $p(\sigma_i | \bmsi_{< i}) = f^{(i)}_\text{out}(\arr{h}^{(i)})$. Therefore, the RNN quantum state is defined by the update function $f^{(i)}_\text{update}$, the output function $f^{(i)}_\text{out}$, and the phase function $\phi$, as illustrated in \Fig{mps-rnn}~(a).
The purpose of introducing $\arr{h}^{(i)}$ instead of working directly with the spin values is the compression of the exponential amount of information in the wave function while keeping a rich expressivity. The trade-off between tractability and expressivity is usually controlled by the complexity of $f^{(i)}_\text{update}$ and $f^{(i)}_{\text{out}}$, and the size of $\arr{h}^{(i)}$.
A notable feature of RNN quantum states is the perfect sampling from the probability distribution of the spins, without the need of Markov chains and therefore avoiding the problem of autocorrelation, which is particularly advantageous when stochastically optimizing the wave function by VMC.

\section{Mapping MPS to RNN}

In the following, we present an exact mapping from an MPS of bond dimension $\chi$ to a specific RNN architecture of memory dimension $\chi$. Accordingly, in this manuscript, we use the term ``bond dimension'' when referring to the dimension of the memory of our RNN architectures to underline the fact that they are conceptually similar.
It is perhaps not completely surprising that a mapping from MPS to RNN exists as MPS has been shown to allow perfect sampling~\cite{ferris2012perfect} and as it achieves an efficient compression of the information of the wave function for many quantum systems.

The MPS ansatz for a wave function $\psi$ is defined by
\begin{equation}
\psi(\bmsi) = \sum_{s_0, \ldots, s_{V} = 0}^{\chi - 1} \prod_{i = 0}^{V - 1} M^{(i)}_{\sigma_i; s_{i + 1}, s_i},
\label{eq:mps-psi}
\end{equation}
where $\arr{M}^{(i)}_{\sigma_i}$ is a complex $\chi \times \chi$ matrix that depends on the spin $\sigma_i$ at the site $i$. Note that we keep the indices $s_0$ and $s_{V}$ at both ends to simplify the discussion.
To rewrite the MPS as an RNN, we identify the intermediate result of the tensor contraction as the memory $\arr{h}^{(i)}$, which satisfies a local update rule
\begin{equation}
\arr{h}^{(i)} = \arr{M}^{(i)}_{\sigma_i}\,\arr{h}^{(i - 1)},
\label{eq:mps-rnn-h}
\end{equation}
where $\arr{h}^{(i)} \in \mathbb{C}^\chi$ is a vector for each site $i$, and implicitly depends on the previous spins $\bmsi_{\le i} = (\sigma_0, \ldots, \sigma_i)$. The boundary condition is $\arr{h}^{(-1)} = (1, \ldots, 1)$. This is useful as the conditional probability of the MPS is then proportional to a positive semi-definite quadratic form of the memory
\begin{equation}
p(\sigma_i | \bmsi_{< i}) \propto [\arr{h}^{(i)}]^\dagger\,\arr{\gamma}^{(i)}\,\arr{h}^{(i)},
\label{eq:mps-rnn-p}
\end{equation}
where the explicit form of $\arr{\gamma}^{(i)}$ as a contraction of the $\arr{M}^{(i)}_\sigma$ matrices is given for completeness in the Supplemental Material~\ref{append:gamma}. The phase of $\psi$ can be obtained from the memory at the last site:
\begin{equation}
\phi(\bmsi) = \arg \sum_s h^{(V - 1)}_s.
\label{eq:mps-rnn-phase}
\end{equation}

We now have all the elements to introduce the ansatz that we refer to as the vanilla MPS-RNN: \Eq{mps-rnn-h} defines the memory update in terms of the variational parameters $\arr{M}^{(i)}_\sigma$, the conditional probability is obtained from \Eq{mps-rnn-p}, and the phase is obtained from \Eq{mps-rnn-phase}. See \Fig{mps-rnn}~(b) for TN diagrammatic illustrations of the memory update and the conditional probability. We elevate each $\arr{\gamma}^{(i)}$ to be a free variational parameter, independent of $\arr{M}^{(i)}_\sigma$. Therefore, while any MPS can be exactly mapped to a vanilla MPS-RNN with the same bond dimension, the opposite is not true, and the freedom of $\arr{\gamma}^{(i)}$ brings additional expressivity. As $\arr{M}^{(i)}_\sigma$ and $\arr{\gamma}^{(i)}$ depend on the spatial position $i$, we can say that we are encoding the spatial dimension into the ``time'' dimension of the RNN.

We slightly generalize the vanilla MPS-RNN architecture to improve its numerical performance by adding the normalization of the memory at each step and including a vectorial term in the memory update, as described in the Supplemental Material~\ref{append:1d-mps-rnn}. We call the resulting ansatz 1D~MPS-RNN, which is the one we use in numerical experiments.

\begin{figure}[t]
\includegraphics[width=0.9\linewidth]{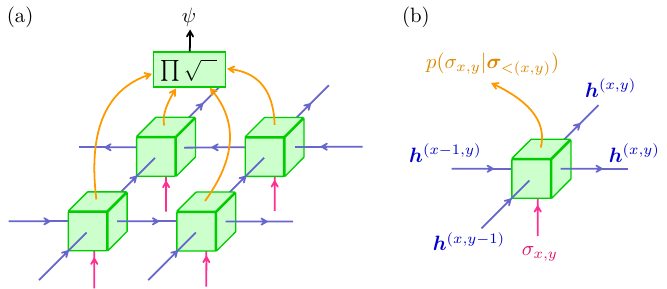}
\caption{
(a) Computational graph of 2D~MPS-RNN and tensor-RNN architectures for square lattice with snake ordering.
(b) Inputs and outputs at each site of 2D~MPS-RNN and tensor-RNN architectures. For each site of the square lattice $(x, y)$, the RNN takes as input the memory vectors $\arr{h}^{(x - 1, y)}$ and $\arr{h}^{(x, y - 1)}$ from previous neighboring sites and the spin value $\sigma_{x, y}$ at the current site, and outputs the updated memory vector $\arr{h}^{(x, y)}$ to be used in future neighboring sites and the conditional probability $p(\sigma_{x, y} | \bmsi_{< (x, y)})$.
}
\label{fig:2d-mps-rnn}
\end{figure}

\section{2D~MPS-RNN}

From the known success of MPS and its exact mapping to 1D~MPS-RNN, we conclude that the latter can represent the ground states of gapped 1D systems. Therefore, the nonlinearity of conventional RNN is not necessary to efficiently represent short-range entanglement in one dimension, as 1D~MPS-RNN only uses a linear memory update. By taking inspiration from the PEPS architecture, we consider a minimal generalization of the architecture to 2D systems, where we seek to efficiently approximate their ground states using linear or multilinear memory updates while keeping the computation time and memory scaling polynomially with the system size.

In the following, for simplicity, we limit our discussion to a square lattice of size $V = L \times L$. As the RNN sequentially outputs the conditional probability of a given spin at each step, we need to define a 1D ordering for the sites on the 2D lattice. We use the ``snake'' ordering as commonly used in MPS~\cite{stoudenmire2012study} and RNN~\cite{hibat2020recurrent} for 2D inputs, which is illustrated in \Fig{2d-mps-rnn}.
A minimal generalization of \Eq{mps-rnn-h} to 2D consists in using a linear combination of the memory components at two previous neighboring sites, and we call the resulting architecture 2D~MPS-RNN. We provide the TN diagram for the memory update of 2D~MPS-RNN in \Fig{tensor-rnn-h}, which is a generalization of \Fig{mps-rnn}~(b). The equations are detailed in the Supplemental Material~\ref{append:tensor-rnn}.

A notable difference between 2D~MPS-RNN and TN is that in the former each memory vector is used in two future neighboring sites. In a TN, on the contrary, each tensor only appears once in the contraction. This reuse of information can lead to a more efficient state compression using fewer parameters~\cite{levine2019quantum}. We remark that while the shallow RAC TN in~\cite{levine2017long} is only a subset of MPS, our vanilla MPS-RNN is a superset of MPS.

The direct information flow between vertically neighboring sites is an apparent advantage of 2D~MPS-RNN over 1D architectures. Thanks to this, the memory at a given site no longer needs to be carried along $O(L)$ memory updates to influence the memory of a vertical neighbor of the 2D lattice. However, because of the linearity of the memory update proposed above, a 2D~MPS-RNN with bond dimension $\chi$ can be exactly simulated by a 1D~MPS-RNN with bond dimension $L\,\chi$, as proved in the Supplemental Material~\ref{append:2d-simulated}. Therefore, 2D~MPS-RNN can reduce the bond dimension $\chi$ at most linearly in $L$ compared to 1D~MPS-RNN rather than exponentially. In particular, its entanglement entropy cannot have an area law because it can scale at most logarithmically with the length $L$ of a horizontal cut at a fixed $\chi$.

\begin{figure}[t]
\includegraphics[width=0.9\linewidth]{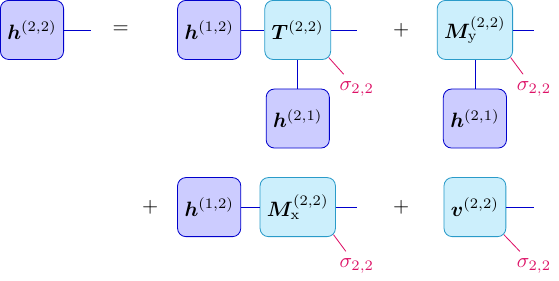}
\caption{TN diagram of the memory update for tensor-RNN. The memory update for 2D~MPS-RNN is obtained for $\arr{T}^{(2, 2)} = \arr{0}$, and the memory update for 1D~MPS-RNN can be derived by further setting $\arr{M}^{(2, 2)}_\text{y} = \arr{0}$.}
\label{fig:tensor-rnn-h}
\end{figure}

\begin{figure*}[t]
\includegraphics[width=\linewidth]{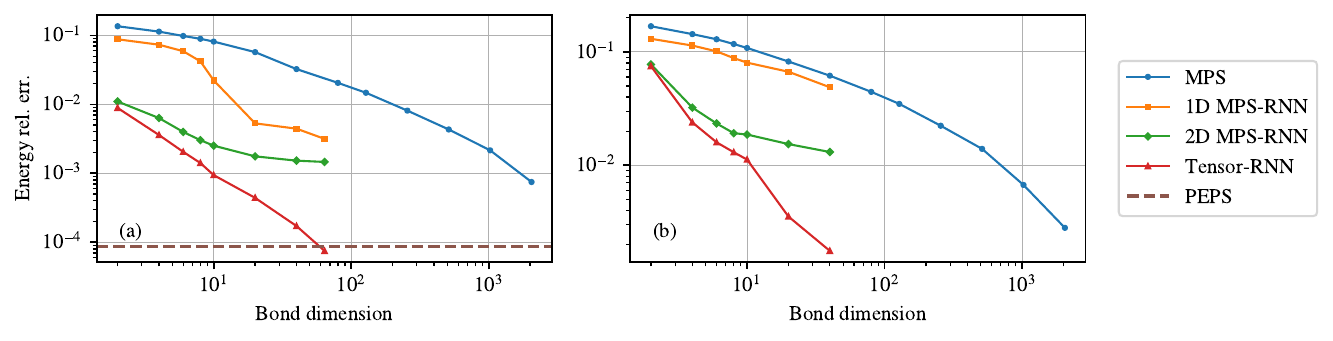}
\caption{Energy relative error of MPS (optimized by DMRG) and our ansätze (optimized by VMC) as a function of the bond dimension $\chi$. The Hamiltonian is AFHM on (a) $10 \times 10$ square lattice and (b) $10 \times 10$ triangular lattice, and the energy is compared to (a) QMC~\cite{sandvik1997finite} and (b) DMRG with $\chi = 4096$. The PEPS result in (a) is taken from~\cite{liu2017gradient} with $\chi = 10$.}
\label{fig:energy}
\end{figure*}

\section{Tensor-RNN}

We are therefore led to consider a nonlinear generalization of the memory update \Eq{mps-rnn-h}, and a minimal choice is a multilinear function of the memories at previous neighboring sites. The resulting memory update is sketched in \Fig{tensor-rnn-h} with the help of a TN diagram and detailed in the Supplemental Material~\ref{append:tensor-rnn}. We call this architecture tensor-RNN. It can represent states with area law of entanglement entropy, and an example of this is constructed in the Supplemental Material~\ref{append:area-law}.

Compared to other methods employing 2D RNN as NQS, such as~\cite{hibat2020recurrent, hibat2021variational}, our ansätze are developed by generalizing TN and therefore retain some advantages of TN, namely (1) their expressivity is controlled by a single hyperparameter, the bond dimension, which avoids the laborious architecture search of RNN; (2) their linear or multilinear architectures are simple enough to enable theoretical analysis tools inspired from TN, such as entanglement entropy; and (3) they provide an effecive way of initialization when used in VMC, which we discuss below.

The ansätze we have discussed form a hierarchy: MPS $\subsetneq$ 1D~MPS-RNN $\subsetneq$ 2D~MPS-RNN $\subsetneq$ tensor-RNN, where an architecture on the right can exactly simulate an architecture on the left using the same bond dimension. We refer to this procedure as \emph{hierarchical initialization}. During the variational optimization, we first use the DMRG algorithm to optimize a MPS, then use the optimized parameters to initialize a 1D~MPS-RNN. After the gradient-based optimization of the 1D~MPS-RNN ansatz, we use it to initialize a 2D~MPS-RNN, which after optimization can be used to initialize a tensor-RNN. This procedure provides reasonable starting points for each optimization and avoids being stuck early in a local minimum with high energy.

\section{Numerical experiments}

To numerically evaluate the performances of our proposed ansätze, we start with the AFHM, defined by the Hamiltonian $\hat{H} = \sum_{\langle i, j \rangle} \hat{\arr{S}}^{(i)} \cdot \hat{\arr{S}}^{(j)}$, where $\langle i, j \rangle$ denotes a pair of nearest neighbors in the lattice. We perform numerical experiments on a $10 \times 10$ square lattice with open boundary conditions (OBC).
We apply the Marshall sign rule~\cite{marshall1955antiferromagnetism} to the Hamiltonian, which makes the ground-state wave function positive. We use the standard VMC method to optimize our ansätze, with automatic differentiation (AD)~\cite{baydin2018automatic} to compute the gradients, perfect sampling~\cite{sharir2020deep}, Adam optimizer~\cite{kingma2014adam}, and the hierarchical initialization described above. The variational energies are compared to the quantum Monte Carlo (QMC) result~\cite{sandvik1997finite}. More information on the numerical implementation can be found in the Supplemental Material~\ref{append:numerical}. \Fig{energy}~(a) shows that we can systematically obtain lower variational energies by increasing the bond dimension $\chi$. We remark that the memory updates of 2D~MPS-RNN and tensor-RNN can reduce the bond dimension by $1$--$2$ orders of magnitude, which confirms our expectation for 2D systems. The significant compression achieved by tensor-RNN over 2D~MPS-RNN shows that the tensorial part of the memory update is crucial to capturing 2D quantum entanglement, as expected from the TN analogy.

Finally, we apply our ansätze on a frustrated system, namely the AFHM on a $10 \times 10$ triangular lattice with OBC. The ground state wave function can no longer be made positive by the Marshall sign rule, and the hierarchical initialization is particularly useful in this case, as it alleviates the difficulty of learning the sign structure of the wave function for frustrated systems~\cite{westerhout2020generalization, bukov2021learning}. \Fig{energy}~(c) shows that the reduction of bond dimension in 2D~MPS-RNN and tensor-RNN is unaffected by the frustration. Moreover, a comparison of square and triangular lattices provides evidence that the multilinear memory update of tensor-RNN is important to capture the sign structure.

\section{Conclusions}

We have shown that an RNN can exactly represent an MPS of bond dimension $\chi$ with a linear memory update, which we call 1D~MPS-RNN. Using analogies with TN and analytical arguments, we have proposed two minimal generalizations of the RNN architecture better suited to capturing 2D quantum entanglement while still using linear or multilinear memory updates. The tensorial version, tensor-RNN, can represent states with the area law of entanglement entropy while keeping the convenient properties of efficient, perfect sampling and wave function evaluation. We have provided numerical evidence for the square AFHM that our ansätze can achieve performance comparable to an MPS of bond dimension higher by orders of magnitude. These results have also been confirmed in the presence of frustration in the case of the triangular AFHM, where we have found that the hierarchical initialization from MPS to tensor-RNN is particularly useful to learn the sign structure of the wave function, which is generally a challenge for NQS.

Several possible future directions can be envisaged. In this work, we have not used one of the essential features of neural networks: nonlinear activation. Future work is needed to understand to what extent nonlinearities help increase the expressivity of representing quantum states and how they affect the optimization of tensorial RNNs.
It has also been shown that deep architectures in modern neural networks can more efficiently produce entanglement~\cite{Sharir22NQSTN}, and our tensorial architectures can be generalized in a multi-layer fashion.
Another promising research direction for the architectures concerns physical symmetries, and we may incorporate the quantum number conservation and the $\mathrm{SU}(2)$ symmetry of MPS~\cite{gong2014plaquette} into our ansätze. With those techniques, we expect tensorial RNNs to find broader applications.

Our code is available at \url{https://github.com/wdphy16/mps-rnn}

\begin{acknowledgments}
We thank Jannes Nys and Or Sharir for their valuable remarks.
Support from the Swiss National Science Foundation is acknowledged under Grant No. 200021\_200336.
\end{acknowledgments}

\clearpage
\widetext

\begin{center}
\textbf{\large Supplemental Material for ``From Tensor Network Quantum States to Tensorial Recurrent Neural Networks''}
\end{center}

\setcounter{section}{0}
\setcounter{equation}{0}
\setcounter{figure}{0}
\setcounter{table}{0}

\renewcommand{\thesection}{S\arabic{section}}
\renewcommand{\theequation}{S\arabic{equation}}
\renewcommand{\thefigure}{S\arabic{figure}}
\renewcommand{\thetable}{S\arabic{table}}

\makeatletter
\c@secnumdepth=3
\makeatother

\section{Conditional probability of MPS} \label{append:gamma}

\begin{figure}[htb]
\includegraphics[width=0.5\linewidth]{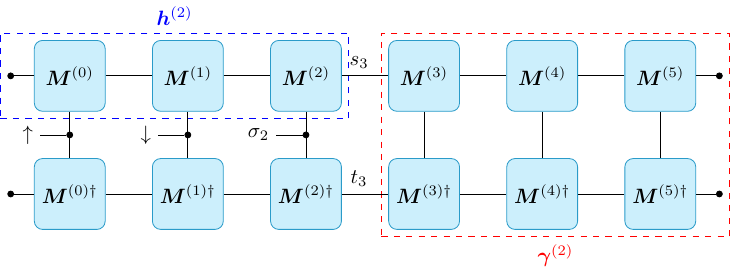}
\caption{Tensor contraction of a MPS with $6$ sites to obtain the unnormalized conditional probability $p(\sigma_2 | \sigma_0 = \spinup, \sigma_1 = \spindown)$.}
\label{fig:mps-cond-prob}
\end{figure}

The construction of conditional probabilities in the vanilla MPS-RNN comes from the fact that in MPS we can efficiently trace out the spins $\bmsi_{> i}$ and specify the spins $\bmsi_{< i}$ to compute the unnormalized conditional probability $p(\sigma_i | \bmsi_{< i})$, as shown in \Fig{mps-cond-prob}, and the normalization is given by
\begin{equation}
\sum_{\sigma \in \{\spinup, \spindown\}} p(\sigma | \bmsi_{< i}) = 1.
\end{equation}
When mapping an MPS to a vanilla MPS-RNN, each coefficient matrix $\arr{\gamma}^{(i)}$ can be computed from the traced out matrices $\{\arr{M}^{(j)}_\sigma | j > i\}$:
\begin{equation}
\gamma^{(i - 1)}_{t_i, s_i} = \sum_{\substack{s_{i + 1}, \ldots, s_V \\ t_{i + 1}, \ldots, t_V \\ \sigma_i, \ldots, \sigma_{V - 1}}}
\prod_{j = i}^{V - 1} [M^{(j)}_{\sigma_j; t_{j + 1}, t_j}]^*\,M^{(j)}_{\sigma_j; s_{j + 1}, s_j}.
\end{equation}

\section{1D~MPS-RNN} \label{append:1d-mps-rnn}

We append a few generalizations to the vanilla MPS-RNN defined by Eqs.~\eqref{eq:mps-rnn-h}, \eqref{eq:mps-rnn-p}, and \eqref{eq:mps-rnn-phase} to ease its practical use in numerical experiments with gradient-based optimization. As each $\arr{\gamma}^{(i)}$ matrix is positive semidefinite, we diagonalize it by $\arr{\gamma}^{(i)} = [\arr{U}^{(i)}]^\dagger\,\diag(\arr{\eta}^{(i)})\,\arr{U}^{(i)}$, where $\arr{U}^{(i)}$ is a unitary matrix and $\arr{\eta}^{(i)}$ is a real and positive vector. We further absorb $\arr{U}^{(i)}$ by $\arr{h}^{(i)} \gets \arr{U}^{(i + 1)}\,\arr{h}^{(i)}$ and $\arr{M}^{(i)}_\sigma \gets \arr{U}^{(i + 1)}\,\arr{M}^{(i)}_\sigma\,[\arr{U}^{(i)}]^\dagger$, so that \Eq{mps-rnn-p} is simplified without loss of generality to
\begin{equation}
p(\sigma_i | \bmsi_{< i}) \propto \sum_{s} \eta^{(i)}_s\,|h^{(i)}_s|^2.
\label{eq:1d-mps-rnn-p}
\end{equation}
During the variational optimization, we regard the parameters $\arr{\eta}^{(i)}$ as independent of $\arr{M}^{(i)}_\sigma$, and we use $\log \eta^{(i)}_s$ as free parameters to ensure the positivity of $\eta^{(i)}_s$. These free parameters give our ansatz richer expressivity than the original MPS.

Based on \Eq{mps-rnn-h}, we add a vector term $\arr{v}^{(i)}_\sigma \in \mathbb{C}^\chi$, which plays the role of a bias in linear layers of NN, and normalize $\arr{h}^{(i)}$ at each step to improve numerical stability:
\begin{gather}
\tilde{\arr{h}}^{(i)}(\sigma_i, \bmsi_{< i}) = \arr{M}^{(i)}_{\sigma_i}\,\arr{h}^{(i - 1)}(\sigma_{i - 1}, \bmsi_{< i - 1}) + \arr{v}^{(i)}_{\sigma_i},
\label{eq:1d-mps-rnn-h} \\
h^{(i)}_s(\sigma_i, \bmsi_{< i}) = \frac{\tilde{h}^{(i)}_s(\sigma_i, \bmsi_{< i})}{\sqrt{\sum_{t, \sigma} \left\lvert \tilde{h}^{(i)}_t(\sigma, \bmsi_{< i}) \right\rvert^2}}.
\label{eq:h-normalize}
\end{gather}
Note that here we explicitly show the dependence of $\arr{h}^{(i)}$ on the spins $\sigma_i$ and $\bmsi_{< i}$. When training the ansatz, not including the vector term $\arr{v}^{(i)}_\sigma$ can cause gradient explosion because there is no bounded nonlinear activation between RNN steps as in conventional RNN architectures to suppress high-order powers of parameters, and the $\arr{v}^{(i)}_\sigma$ term helps regularize the gradients.
For a trained ansatz, we can always absorb $\arr{v}^{(i)}_\sigma$ into an extra row of $\arr{M}^{(i)}_\sigma$, and the normalization does not affect the conditional probability. \Eq{1d-mps-rnn-h} is also a special case of \Eq{tensor-rnn-h} for tensor-RNN with $\arr{T}^{(x, y)}_\sigma = \arr{0}$, $\arr{M}^{(x, y)}_{\text{y}; \sigma} = \arr{0}$.

We also obtain the phase from a generalization of \Eq{mps-rnn-phase}:
\begin{equation}
\phi(\bmsi) = \sum_i \phi^{(i)}, \quad
\phi^{(i)} = \arg\left( [\arr{w}^{(i)}_{\sigma_i}]^\rmT \arr{h}^{(i)} + c^{(i)}_{\sigma_i} \right),
\label{eq:1d-mps-rnn-phase}
\end{equation}
where $\arr{w}^{(i)}_\sigma \in \mathbb{C}^\chi$ and $c^{(i)}_\sigma \in \mathbb{C}$ are new variational parameters. There is a local phase at each RNN step, which directly contributes to the global phase and receives gradient information from the loss. This representation of the phase is particularly useful when optimizing on frustrated systems, although the physical correspondence of the local phases remains an open question. We call the above architecture 1D~MPS-RNN, which is defined by Eqs.~\eqref{eq:1d-mps-rnn-h}, \eqref{eq:1d-mps-rnn-p}, and \eqref{eq:1d-mps-rnn-phase}.

In the simplest form of commonly used RNN architectures in machine learning, the recurrence equation is defined by the following instead of \Eq{1d-mps-rnn-h}:
\begin{equation}
\arr{h}^{(i)}(\sigma_i, \bmsi_{< i}) = f(\arr{M} \arr{h}^{(i - 1)} + \sigma_i \arr{w} + \arr{v}),
\end{equation}
where $\arr{M}$ is a matrix of parameters, $\arr{w}$ and $\arr{v}$ are vectors of parameters, and they are independent of the site index $i$ or the input $\sigma_i$. The function $f$ is a nonlinear activation function, usually the sigmoid function. More advanced recurrence equations, such as long short-term memory (LSTM)~\cite{hochreiter1997long} and gated recurrent unit (GRU)~\cite{cho2014properties}, have been proposed to better represent long-term correlations. In comparison, we use site- and input-dependent parameters such as $\arr{M}^{(i)}_\sigma$ and $\arr{h}^{(i)}_\sigma$ to create the exact mapping between MPS and RNN, and we do not use the nonlinear activation.

\section{Memory update for 2D~MPS-RNN and tensor-RNN} \label{append:tensor-rnn}

As illustrated in \Fig{tensor-rnn-h}, the memory update equation for tensor-RNN is
\begin{align}
\tilde{h}^{(x, y)}_s = &\phantom{{}+{}} \sum_{t, u} T^{(x, y)}_{\sigma_{x, y}; s, t, u} h^{(x \pm 1, y)}_t h^{(x, y - 1)}_u \nonumber \\
&+ \sum_t M^{(x, y)}_{\text{x}; \sigma_{x, y}; s, t} h^{(x \pm 1, y)}_t
+ \sum_t M^{(x, y)}_{\text{y}; \sigma_{x, y}; s, t} h^{(x, y - 1)}_t \nonumber \\
&+ v^{(x, y)}_{\sigma_{x, y}; s}
\label{eq:tensor-rnn-h}
\end{align}
for lattice indices $x, y \in \{0, \ldots, L - 1\}$, where $\arr{T}^{(x, y)}_{\sigma_{x, y}}$ is a complex tensor of dimensions $\chi \times \chi \times \chi$ for each site $(x, y)$ and depends on the spin $\sigma_{x, y}$. In the matrix terms, the subscripts x and y (in upright font) denote two matrices at the same site and acting on different input memory vectors. The sign $\pm$ takes $-$ when $y$ is even and $+$ when $y$ is odd, as defined by the snake ordering and shown in \Fig{2d-mps-rnn}. The boundary conditions are $\arr{h}^{(-1, 0)} = \arr{1}$ and $\arr{h}^{(-1, y)} = \arr{h}^{(L, y)} = \arr{h}^{(x, -1)} = \arr{0}$ for all $x$ and for $y > 0$. We call this architecture tensor-RNN, which is defined by Eqs.~\eqref{eq:tensor-rnn-h}, \eqref{eq:1d-mps-rnn-p}, and \eqref{eq:1d-mps-rnn-phase}. The 2D~MPS-RNN is a special case of tensor-RNN with $\arr{T}^{(x, y)}_\sigma = \arr{0}$.

To discuss the computational complexity of tensor-RNN, we focus on one RNN step and use simplified notations in \Eq{tensor-rnn-h}:
\begin{equation}
\tilde{h}^{(1)}_s = \sum_{t, u} T_{s, t, u} h^{(2)}_t h^{(3)}_u + \sum_t M^{(\text{x})}_{s, t} h^{(2)}_t + \sum_t M^{(\text{y})}_{s, t} h^{(3)}_t + v_s.
\end{equation}
Evaluating the memory update takes $O(\chi^3)$ time in each RNN step, which is dominated by the tensor term $\sum_{t, u} T_{s, t, u} h^{(2)}_t h^{(3)}_u$ for all $s$, and the evaluations of Eqs.~\eqref{eq:1d-mps-rnn-p} and \eqref{eq:1d-mps-rnn-phase} take a lower order of time compared to \Eq{tensor-rnn-h}. It takes $O(V \chi^3)$ time when running through $V$ RNN steps to evaluate the wave function, where $V = L^2$ is the number of lattice sites. This is exponentially lower than the time to exactly contract a PEPS, and polynomially lower than typical methods of approximate contraction with SVD truncation. For example, the method in~\cite{liu2017gradient} uses $O(\chi^6)$ time. Without the tensor term, the time complexity becomes $O(V \chi^2)$ for 1D and 2D~MPS-RNN.

The static memory (used to store the parameters) is $O(V \chi^2)$ for 1D and 2D~MPS-RNN, and $O(V \chi^3)$ for tensor-RNN. The dynamic memory (additionally used when evaluating the wave function) is $O(\chi)$ for 1D~MPS-RNN as we only need to store the memory of the current site, and $O(L \chi)$ for 2D~MPS-RNN and tensor-RNN as we also need to store the memories in the last row. This is also exponentially lower than exactly contracting a PEPS, and in practice we have found that the bottleneck of scaling up our ansätze is usually time rather than memory.

When computing the gradients of the variational energy w.r.t.\ the parameters of tensor-RNN, following the derivative rule of tensor products, we have
\begin{equation}
\pdv{\tilde{h}^{(1)}_s}{T_{s', t, u}} = \delta_{s, s'} h^{(2)}_t h^{(3)}_u, \quad
\pdv{\tilde{h}^{(1)}_s}{M^{(\text{x})}_{s', t}} = \delta_{s, s'} h^{(2)}_t, \quad
\pdv{\tilde{h}^{(1)}_s}{M^{(\text{y})}_{s', t}} = \delta_{s, s'} h^{(3)}_t, \quad
\pdv{\tilde{h}^{(1)}_s}{v_{s'}} = \delta_{s, s'}.
\end{equation}
Therefore, computing the gradients also takes $O(\chi^3)$ time in each RNN step, which is dominated by the tensor term $\pdv{\tilde{h}^{(1)}_s}{T_{s', t, u}}$ for all $s, t, u$, and it becomes $O(V \chi^3)$ when running through $V$ RNN steps. As the forward evaluation of the wave function also takes $O(V \chi^3)$ time, we conclude that the time complexity of an optimization step is $O(V \chi^3)$. Without the tensor term, it becomes $O(V \chi^2)$ for 1D and 2D~MPS-RNN. The dynamic memory for an optimization step is $O(V \chi)$ for all our ansätze, as we need to store the memories at all sites to compute the gradients using backpropagation.

\section{2D~MPS-RNN simulated by 1D~MPS-RNN} \label{append:2d-simulated}

In the following, we show that a 2D~MPS-RNN with bond dimension $\chi$ can be exactly simulated by a 1D~MPS-RNN with bond dimension $L \chi$. The 1D~MPS-RNN uses the same snake ordering on the 2D lattice, but there is only one matrix term at each site. The idea is that the memory components in each row of the 2D~MPS-RNN are linear combinations of those in the previous row.

Without loss of generality, we assume $y$ is even and $y > 0$. Before updating along the $y$-th row in the 1D~MPS-RNN, the memory $\arr{h}'^{(0, y - 1)}$ is a concatenation of the memories $\arr{h}^{(i, y - 1)}$ of the 2D~MPS-RNN in the previous row:
\begin{equation}
h'^{(0, y - 1)}_{i \chi + s} = h^{(i, y - 1)}_s
\end{equation}
for $i \in \{0, \ldots, L - 1\}$ and $s \in \{0, \ldots, \chi - 1\}$.
As we update along the $y$-th row, the components in $\arr{h}'^{(x, y)}$ are gradually replaced by the memories of the 2D~MPS-RNN in the current row:
\begin{equation}
h'^{(x, y)}_{i \chi + s} = \begin{cases}
h^{(i, y)}_s & \text{if $i \le x$}, \\
h^{(i, y - 1)}_s & \text{if $i > x$}.
\end{cases}
\end{equation}
The corresponding transfer matrix is defined by
\begin{equation}
M'^{(x, y)}_{\sigma; i \chi + s, j \chi + t} = \begin{cases}
M^{(x, y)}_{\text{y}; \sigma; s, t} & \text{if $i = j = x$}, \\
M^{(x, y)}_{\text{x}; \sigma; s, t} & \text{if $i = x - 1$ and $j = x$}, \\
\delta_{s, t} & \text{if $i = j \neq x$}, \\
0 & \text{otherwise}.
\end{cases}
\end{equation}

\section{Tensor-RNN can represent a quantum state with an area law of entanglement entropy} \label{append:area-law}

In the following, we construct a tensor-RNN that has an exact area law of entanglement entropy for any region that follows the snake ordering and has at least one row. The idea is that we can always have $L$ independent memory vectors on the boundary of the region, and the wave function \Eq{ln-psi} is a nontrivial multilinear function of them.

\begin{figure}[t]
\includegraphics[width=0.3\linewidth]{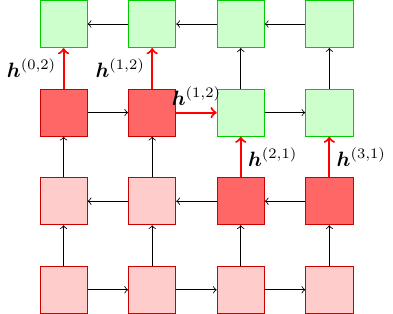}
\caption{Sketch of the LN tensor-RNN architecture. Red sites are in a region following the snake ordering, for which we compute the reduced density matrix $\arr{\rho}^{\le (1, 2)}$. The dark red sites are on the boundary $\calB(1, 2)$, and the red edges are the memories $\arr{h}^{(x, y)}$ on the boundary. Green sites are in the traced-out region.}
\label{fig:ln-tensor-rnn}
\end{figure}

We start by introducing a new type of variational ansatz, which we call late normalizing (LN) tensor-RNN. We use the same update rule as \Eq{tensor-rnn-h} with the snake ordering and an even $L$:
\begin{align}
h^{(x, y)}_s = &\phantom{{}+{}} \sum_{t, u} T^{(x, y)}_{\sigma_{x, y}; s, t, u} h^{(x \pm 1, y)}_t h^{(x, y - 1)}_u \nonumber \\
&+ \sum_t M^{(x, y)}_{\text{x}; \sigma_{x, y}; s, t} h^{(x \pm 1, y)}_t
+ \sum_t M^{(x, y)}_{\text{y}; \sigma_{x, y}; s, t} h^{(x, y - 1)}_t \nonumber \\
&+ v^{(x, y)}_{\sigma_{x, y}; s},
\end{align}
except that we do not apply the normalization \Eq{h-normalize} or compute the conditional probability \Eq{1d-mps-rnn-p} at each step, but only take the $0$-th component of the memory vector at the last step to be the wave function:
\begin{equation}
\tilde{\psi}_\text{LN}(\bmsi) = h^{(0, L - 1)}_0,
\end{equation}
and normalize it afterwards:
\begin{equation}
\psi_\text{LN}(\bmsi) = \frac{\tilde{\psi}_\text{LN}(\bmsi)}{\sqrt{\sum_{\bmsi'} \left\lvert \tilde{\psi}_\text{LN}(\bmsi') \right\rvert^2}}.
\end{equation}
We remark that a general LN tensor-RNN cannot be represented by a tensor-RNN or a TN. Despite that, it is useful for our theoretical analysis, as it allows us to obtain analytical expressions for the entanglement entropy without worrying about the normalization. In the following, we construct a particular LN tensor-RNN such that $\tilde{\psi}_\text{LN}(\bmsi)$ is always a multilinear function of the memories on the boundary, as illustrated in \Fig{ln-tensor-rnn}.

We consider a LN tensor-RNN with $\chi = 2$. In the last row $y = L - 1$ of the lattice, the parameters are defined as
\begin{equation}
T^{(x, L - 1)}_{\sigma; s, t, u} = \begin{cases}
\delta_{s, 0}\,\delta_{t, 0}\,\delta_{u, \sigma} & \text{if $x < L - 1$}, \\
0 & \text{if $x = L - 1$},
\end{cases} \quad
\arr{M}^{(x, L - 1)}_{\text{x}; \sigma} = \arr{0}, \quad
M^{(x, L - 1)}_{\text{y}; \sigma; s, t} = \begin{cases}
0 & \text{if $x < L - 1$}, \\
\delta_{s, 0}\,\delta_{t, \sigma} & \text{if $x = L - 1$},
\end{cases} \quad
\arr{v}^{(x, L - 1)}_\sigma = \arr{0},
\end{equation}
where in $\delta_{u, \sigma}$ we represent $\sigma$ by $\{0, 1\}$. We can check that $\tilde{\psi}_\text{LN}(\bmsi)$ is a multilinear function of the memories on the boundary:
\begin{align}
\tilde{\psi}_\text{LN}(\bmsi) = h^{(0, L - 1)}_0 &= h^{(x + 1, L - 1)}_0 \prod_{x' = 0}^x h^{(x', L - 2)}_{\sigma_{x', L - 1}} \quad \text{for any $x < L - 1$} \\
&= \prod_{x' = 0}^{L - 1} h^{(x', L - 2)}_{\sigma_{x', L - 1}}.
\end{align}
In other rows $y < L - 1$, the parameters are defined as
\begin{equation}
\arr{T}^{(x, y)}_\sigma = \arr{0}, \quad
\arr{M}^{(x, y)}_{\text{x}; \sigma} = \arr{0}, \quad
M^{(x, y)}_{\text{y}; \sigma; s, t} = \begin{cases}
0 & \text{if $y = 0$}, \\
\delta_{s, t} & \text{if $y > 0$},
\end{cases} \quad
v^{(x, y)}_{\sigma; s} = \begin{cases}
\delta_{s, \sigma} & \text{if $y = 0$}, \\
0 & \text{if $y > 0$}.
\end{cases}
\end{equation}
For any $x$ and $0 < y < L - 1$, assuming an even $y$ without loss of generality, we can check that
\begin{gather}
\tilde{\psi}_\text{LN}(\bmsi) = \prod_{x' = 0}^{x - 1} h^{(x', y)}_{\sigma_{x', L - 1}} \prod_{x' = x}^{L - 1} h^{(x', y - 1)}_{\sigma_{x', L - 1}}, \\
h^{(x, y)}_s = \delta_{s, \sigma_{x, 0}}.
\end{gather}
Therefore, we always have $L$ memory vectors on the boundary, each has length $\chi = 2$ and only depends on the spin at the first row and the same column $\sigma_{x, 0}$.
We note that although some $\arr{h}^{(x, y)}$ can appear twice on the boundary, like $\arr{h}^{(1, 2)}$ in \Fig{ln-tensor-rnn}, they still appear only once in $\tilde{\psi}_\text{LN}(\bmsi)$ because of the particular choice of parameters. Eventually, $\tilde{\psi}_\text{LN}(\bmsi)$ can be simplified to
\begin{equation}
\tilde{\psi}_\text{LN}(\bmsi) = \prod_{x = 0}^{L - 1} \delta_{\sigma_{x, 0}, \sigma_{x, L - 1}}.
\label{eq:ln-psi}
\end{equation}

We denote by $\arr{\rho}^{\le (x, y)}$ the reduced density matrix of the system obtained by tracing out $\bmsi_{> (x, y)}$, where $\bmsi_{> (x, y)}$ is the set of all spins $\sigma_{x', y'}$ whose coordinates $(x', y')$ come after $(x, y)$ in the snake ordering of the lattice. Using the wave function in \Eq{ln-psi}, the unnormalized reduced density matrix is
\begin{equation}
\tilde{\rho}^{\le (x, y)}\left( \bmsi_{\le (x, y)}, \bmsi'_{\le (x, y)} \right) = \sum_{\bmsi_{> (x, y)}} \tilde{\psi}_\text{LN}\left( \bmsi_{\le (x, y)}, \bmsi_{> (x, y)} \right) \tilde{\psi}^*_\text{LN}\left( \bmsi'_{\le (x, y)}, \bmsi_{> (x, y)} \right) = \prod_{x = 0}^{L - 1} \delta_{\sigma_{x, 0}, \sigma'_{x, 0}}
\end{equation}
for $0 < y < L - 1$, which shows the area law of the entanglement entropy
\begin{equation}
S = L \log 2.
\end{equation}

Finally, we show that this LN tensor-RNN can be represented by a tensor-RNN. As long as $\tilde{\psi}_\text{LN}(\bmsi)$ is a multilinear function of the memories on the boundary, $\arr{\rho}^{\le (x, y)}$ can be written as a multi-quadratic function of them, analogously to \Eq{mps-rnn-p}:
\begin{equation}
\rho^{\le (x, y)}\left( \bmsi_{\le (x, y)}, \bmsi'_{\le (x, y)} \right) = \sum_{\arr{s}_{\calB(x, y)}, \arr{s}'_{\calB(x, y)}} \calH^{(x, y)}_{\arr{s}_{\calB(x, y)}}\left( \bmsi_{\le (x, y)} \right) \gamma^{> (x, y)}_{\arr{s}_{\calB(x, y)}, \arr{s}'_{\calB(x, y)}}\left( \bmsi_{\le (x, y)}, \bmsi'_{\le (x, y)} \right) \left[ \calH^{(x, y)}_{\arr{s}'_{\calB(x, y)}}\left( \bmsi'_{\le (x, y)} \right) \right]^*,
\label{eq:boundary-memory-rho}
\end{equation}
where $\calB(x, y)$ is the boundary of the region, $\arr{s}_{\calB(x, y)} = \{s_{x', y'} | (x', y') \in \calB(x, y)\}$ are the memory indices on the boundary,
\begin{equation}
\calH^{(x, y)}_{\arr{s}_{\calB(x, y)}}\left( \bmsi_{\le (x, y)} \right) = \prod_{(x', y') \in \calB(x, y)} h^{(x', y')}_{s_{x', y'}}\left( \bmsi_{\le (x, y)} \right)
\end{equation}
is the direct product of all memory vectors on the boundary, and $\arr{\gamma}^{> (x, y)}$ is a tensor. If we can also write $\arr{\gamma}^{> (x, y)}$ as a direct product over sites:
\begin{equation}
\gamma^{> (x, y)}_{\arr{s}_{\calB(x, y)}, \arr{s}'_{\calB(x, y)}} = \prod_{(x', y') \in \calB(x, y)} \gamma^{(x', y')}_{s_{x', y'}, s'_{x', y'}},
\end{equation}
and it does not depend on the spins $\bmsi_{\le (x, y)}, \bmsi'_{\le (x, y)}$, then we can represent the LN tensor-RNN by a tensor-RNN. With our particular choice of parameters, we simply have $\gamma^{(x, y)}_{s_{x, y}, s'_{x, y}} = \delta_{s_{x, y}, s'_{x, y}}$. Along with the above definitions of $\arr{T}^{(x, y)}_\sigma, \arr{M}^{(x, y)}_{\text{x}; \sigma}, \arr{M}^{(x, y)}_{\text{y}; \sigma}, \arr{v}^{(x, y)}_\sigma$, we define the remaining parameters of the tensor-RNN as
\begin{gather}
\arr{\eta}^{(x, y)} = \arr{1} \quad \text{for $y < L - 1$}, \quad
\eta^{(x, L - 1)}_s = \delta_{s, 0}, \\
w^{(x, y)}_{\sigma; s} = \begin{cases}
\delta_{s, 0} & \text{if $x = 0$ and $y = L - 1$}, \\
0 & \text{otherwise},
\end{cases} \quad
c^{(x, y)}_\sigma = \begin{cases}
0 & \text{if $x = 0$ and $y = L - 1$}, \\
1 & \text{otherwise}.
\end{cases}
\end{gather}

\section{Compressed tensor-RNN} \label{append:compress}

The tensor term in \Eq{tensor-rnn-h} introduces an $O(\chi^3)$-growing time and memory consumption when evaluating the tensor-RNN, which becomes a challenge when $\chi$ gets large. To avoid that, we apply a Tucker decomposition~\cite{tucker1966some} of the tensor:
\begin{equation}
T^{(x, y)}_{\sigma; s, t, u} = \sum_{s', t', u'} K^{(x, y)}_{\sigma; s', t', u'} U^{(x, y)}_{\text{o}; \sigma; s, s'}  U^{(x, y)}_{\text{x}; \sigma; t, t'} U^{(x, y)}_{\text{y}; \sigma; u, u'},
\label{eq:tensor-decomp}
\end{equation}
where $\arr{K}^{(x, y)}_\sigma$ is a smaller tensor of dimensions $\chi' \times \chi' \times \chi'$, and we heuristically choose $\chi' = \chi^{2/3}$ (rounded up to an integer) so that the additional time and memory consumption introduced by the tensor term is limited to $O(\chi^2)$. During the variational optimization, $\arr{K}^{(x, y)}_\sigma$, $\arr{U}^{(x, y)}_{\text{o}; \sigma}$, $\arr{U}^{(x, y)}_{\text{x}; \sigma}$, and $\arr{U}^{(x, y)}_{\text{y}; \sigma}$ are all regarded free parameters. We call this architecture compressed tensor-RNN. In the numerical results, we have found that compressed tensor-RNN intuitively produces a variational energy between 2D~MPS-RNN and tensor-RNN, whether with a fixed $\chi$ or a fixed number of parameters.

\section{Details of numerical experiments} \label{append:numerical}

The RNN architectures are implemented using \texttt{NetKet}~\cite{netket2, netket3}. In the numerical experiments on the square AFHM, as we know the grounds states are real and positive, we discard the parameters $\arr{w}^{(i)}_\sigma$ and $c^{(i)}_\sigma$ and fix the phase to $0$. For the triangular AFHM, the ground state is no longer positive and we indeed need those parameters for the phase. The triangular lattice is defined by adding a diagonal edge to each plaquette in the square lattice.

During the hierarchical initialization, for the $\arr{M}^{(x, y)}_{\text{x}; \sigma}$ and $\arr{M}^{(x, y)}_{\text{y}; \sigma}$ in the 2D~MPS-RNN that are undefined in the corresponding 1D~MPS-RNN, and $\arr{T}^{(x, y)}_\sigma$ in the tensor-RNN, we initialize them using a element-wise Gaussian distribution with standard deviation $10^{-7}$ to avoid degeneracy of features.

During the training, we use Adam optimizer with batch size $1,024$ for all reported results, and we have found that the stochastic reconfiguration (SR) and the plain stochastic gradient descent (SGD) generally produce higher variational energies within the same number of training steps or wall-clock time. Gradient clipping with global norm $1$ is applied to alleviate the energy spikes. After the training, each reported variational energy is estimated from $10^6$ samples, and its statistical error is too small to be visible in the plots.

The results are particularly sensitive to the learning rate, because of the instability discussed in Section~\ref{append:1d-mps-rnn}. For bond dimensions $\chi \le 10$, we use learning rate $10^{-2}$ for $10,000$ steps, $10^{-3}$ for another $10,000$ steps, and finally $10^{-4}$ for $20,000$ steps. We manually reduce the learning rate by a factor of $10$ if we find the variational energy diverges in the middle of training. For bond dimensions $\chi > 10$, we start from learning rate $10^{-3}$. The training of a tensor-RNN with $\chi = 64$ on the $10 \times 10$ triangular AFHM takes about $2$ days on an Nvidia V100 GPU.

We have computed all the DMRG results in \Fig{energy}, except for $\chi \ge 1,024$ on the $20 \times 20$ lattice. The DMRG algorithm is implemented using \texttt{ITensors.jl}~\cite{itensor}, with the snake ordering, $20$ sweeps, eigenvalue cut-off threshold $10^{-12}$, and noises decaying exponentially from $10^{-3}$ to $10^{-12}$ in the first $10$ sweeps.

\section{Energy error as function of the number of parameters} \label{append:E-param}

\begin{figure}[htb]
\includegraphics[width=\linewidth]{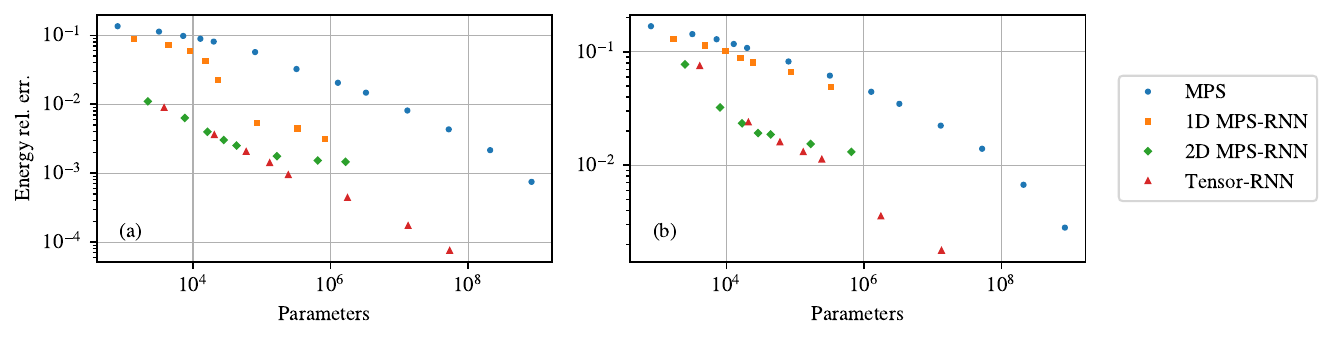}
\caption{Energy relative error in \Fig{energy} as a function of the number of parameters. Our ansätze only need $2$ to $3$ orders of magnitude fewer parameters than MPS to obtain the same energy error. At a fixed number of parameters, there is a clear difference in the energy error between the 1D and the 2D ansätze, which shows that the 2D ansätze can more efficiently compress the approximated ground state.}
\label{fig:E-param}
\end{figure}

\clearpage

\section{Correlations of spins}

\begin{figure}[htb]
\includegraphics[width=0.9\linewidth]{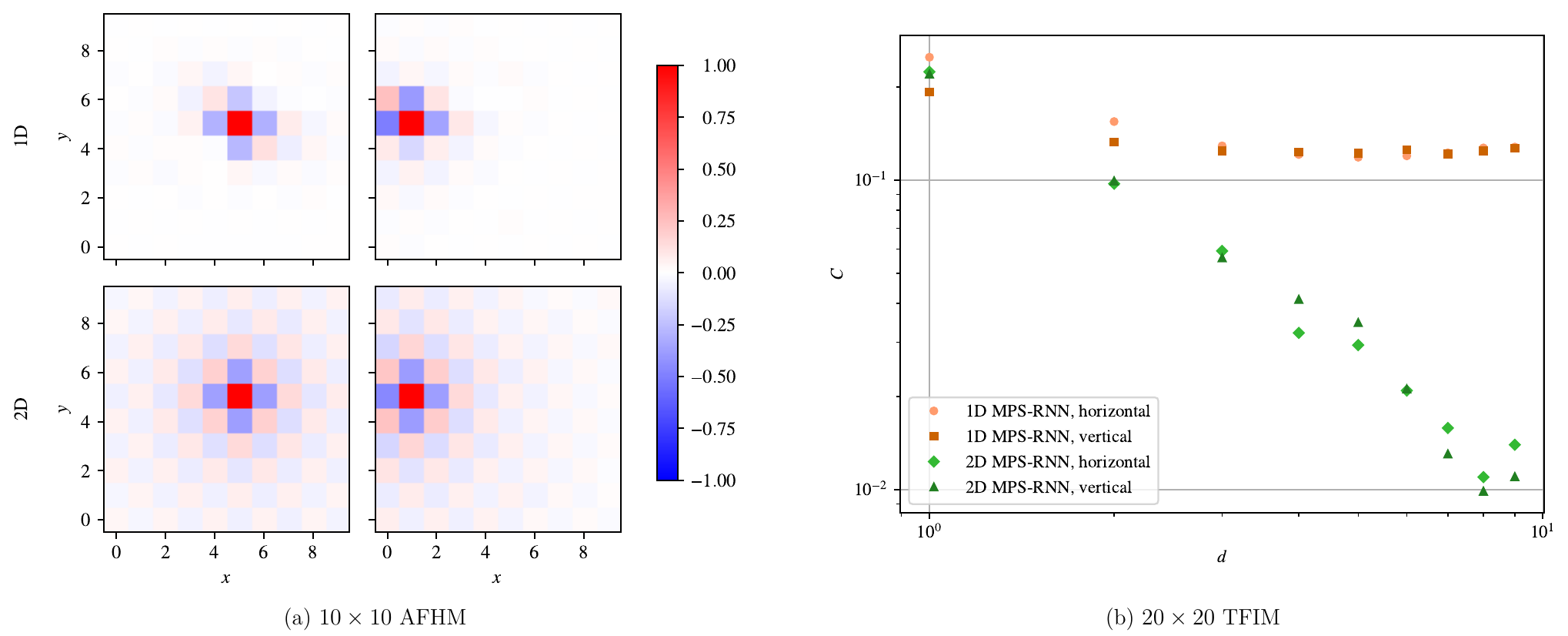}
\caption{
(a) Qualitative evaluation of the correlations of all spins with the spin $\sigma_{5, 5}$ (left column) and $\sigma_{1, 5}$ (right column), in a 1D~MPS-RNN (top row) and a 2D~MPS-RNN (bottom row) with $\chi = 10$ trained on the $10 \times 10$ square AFHM. The 2D~MPS-RNN produces more isotropic correlations, especially near the ends of rows, thanks to the vertical connections.
(b) Quantitative evaluation of the correlations $C$ of the spins $\sigma_{d, 10}$ with $\sigma_{0, 10}$ (horizontal) and the spins $\sigma_{10, d}$ with $\sigma_{10, 0}$ (vertical), in a 1D~MPS-RNN and a 2D~MPS-RNN with $\chi = 16$ trained on the $20 \times 20$ transverse field Ising model at the critical point. The 2D~MPS-RNN captures the algebraic decay of the correlations in both directions, while the 1D one fails to do so.
}
\label{fig:corr}
\end{figure}

\section{Contributions of tensor, matrix, and vector terms to the memory}

\begin{figure}[htb]
\includegraphics[width=0.9\linewidth]{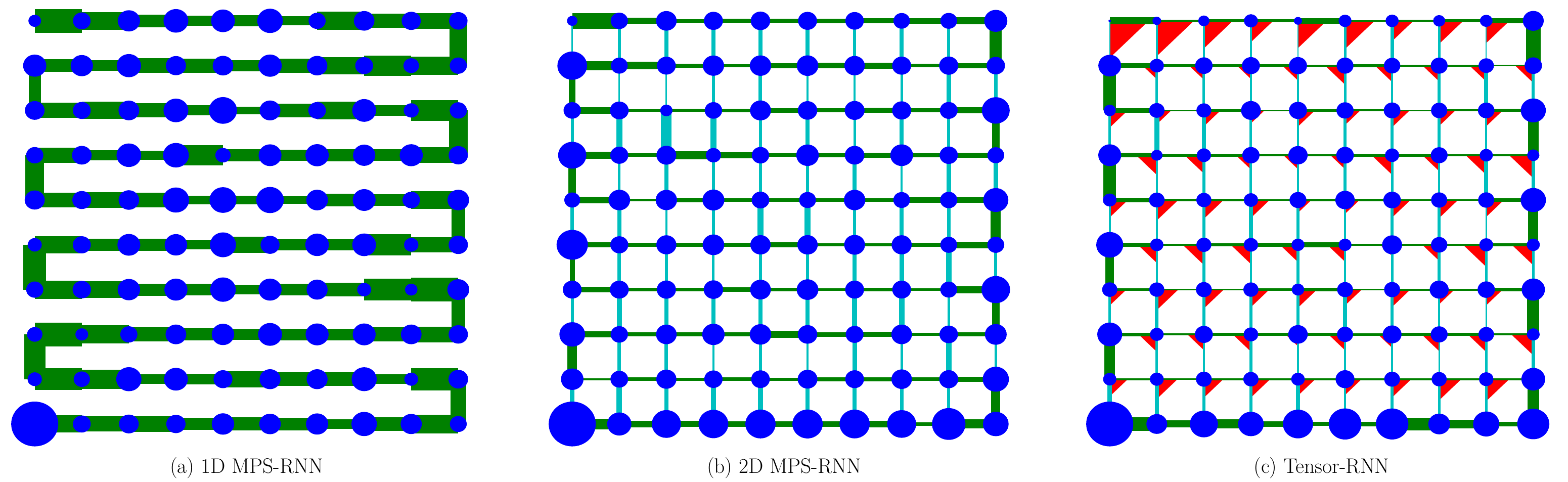}
\caption{Contributions of the tensor term $\arr{T}^{(x, y)}_{\sigma_{x, y}} \arr{h}^{(x \pm 1, y)} \otimes \arr{h}^{(x, y - 1)}$ (areas of red triangles), the matrix terms $\arr{M}^{(x, y)}_{\text{x}; \sigma_{x, y}}\,\arr{h}^{(x \pm 1, y)}$ and $\arr{M}^{(x, y)}_{\text{y}; \sigma_{x, y}}\,\arr{h}^{(x, y - 1)}$ (areas of rectangles), and the vector term $\arr{v}^{(x, y)}_{\sigma_{x, y}}$ (areas of blue circles) in \Eq{tensor-rnn-h} at each site $(x, y)$, in our ansätze with $\chi = 10$ trained on the $10 \times 10$ square AFHM. The green rectangles are the matrix terms along the snake ordering, and the cyan ones are not. The magnitude of each term $\arr{a}$ is defined by the norm $\sqrt{\sum_s \eta^{(x, y)}_s |a_s|^2}$, then averaged over the configurations $\bmsi \sim |\psi(\bmsi)|^2$. At each site, we normalize the four magnitudes to see their relative contributions to the memory at that site. In 1D~MPS-RNN, the matrix term becomes larger near the ends of each row, because it needs to carry all information in the current row to the next row. In 2D~MPS-RNN that is not the case, thanks to the vertical connections. In tensor-RNN, the tensor terms in the last row are particularly large, because they are essential for expressing the wave function as a multilinear function of the memories and representing the area law of entanglement entropy, as shown by the example in Section~\ref{append:area-law}.}
\label{fig:h-components}
\end{figure}

\section{Effect of hierarchical initialization}

\begin{figure}[htb]
\includegraphics[width=0.6\linewidth]{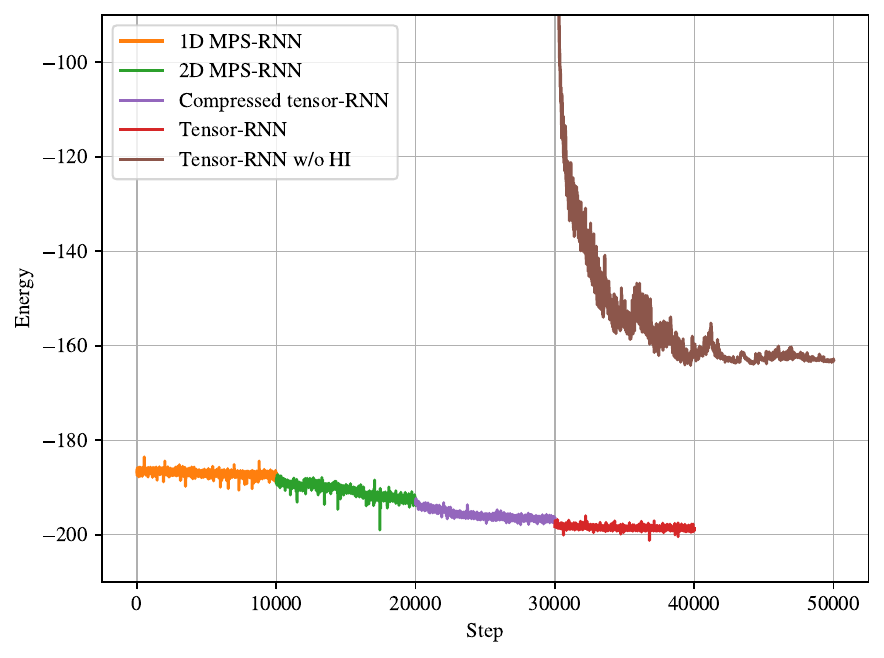}
\caption{Energy in a typical optimization procedure with hierarchical initialization (HI) through 1D~MPS-RNN, 2D~MPS-RNN, compressed tensor-RNN (see Sec.~\ref{append:compress}), and tensor-RNN, compared to a tensor-RNN with random initialization instead of HI. The Hamiltonian is the $10 \times 10$ triangular AFHM, and all ansätze have bond dimension $\chi = 16$. The HI procedure starts at the energy of DMRG, and we can clearly see that each stage of HI lowers the energy as new variational parameters are introduced, while the wave function without HI gets stuck at a high energy.}
\label{fig:hierarchical-training}
\end{figure}


\input{main.bbl}

\end{document}

%% file: main.bbl
%